# Ethics lines and Machine learning: a design and simulation of an Association Rules Algorithm for exploiting the data

Patrici Calvo & Rebeca Egea-Moreno
Universitat Jaume I

**Abstract:** Data mining techniques offer great opportunities for developing ethics lines – tools for communication, participation and innovation whose main aim is to ensure improvements and compliance with the values, conduct and commitments making up the code of ethics. The aim of this study is to suggest a process for exploiting the data generated by the data generated and collected from an ethics line by extracting rules of association and applying the Apriori algorithm. This makes it possible to identify anomalies and behaviour patterns requiring action to review, correct, promote or expand them, as appropriate. Finally, I offer a simulated application of the Apriori algorithm, supplying it with synthetic data to find out its potential, strengths and limitations.

## 0. INTRODUCTION

The sudden emergence of disruptive techniques and technologies linked to the synergistic convergence of the Internet of Things (IoT), Big Data and Artificial Intelligence (AI) and their application in fields like policing, economics, health and entertainment is shaping a new model of society based on the digital hyperconnectivity of all reality, the datafication of social action and the algorithmisation of decision-making processes. The underlying socio-political, economic and cultural context offers new horizons of opportunities for the different areas of human activity. As has been seen over the last few years, there is clear evidence that the application of digital techniques and technologies offers competitive, strategic and evolutionary advantages in terms of sustainability, predictability, speed, exhaustiveness, extensibility, capacity, completeness, consistency, efficiency, ratification, precision and detection, among other things.

One of the areas where the application and use of digital techniques and technologies arouses the greatest interest is in managing, monitoring and complying with ethics in practice. This particularly applies to the innovative development of ethics lines. These are among the communication and participation tools used by institutions, organisations and companies of different sizes to prevent underlying conflict and improve corporate culture while increasing knowledge of the level of compliance with the commitments made and conduct expected by all agents involved.

Among the digital techniques and technologies that could be applied in developing ethics lines, the most important are machine learning, blockchain and data mining (Calvo, 2021). Focusing on the last of these, data mining, this study is intended is to look in depth at the possibilities of developing ethics lines by applying the Apriori algorithm to exploit the data[1] compiled and generated. To do this, firstly I will critically consider the origins and development of ethics lines. Secondly, I will suggest ethics lines that measure up to current expectations. Thirdly, I will look in depth at the development of ethics lines using data mining techniques. Fourthly and finally, a process for exploiting the data generated and collected by the ethics lines is suggested, alongside a simulation of their application using the Apriori algorithm to extract rules of association (Agrawal et al., 1993).

1. **THE EMERGENCE AND DEVELOPMENT OF ETHICS LINES**

Ethics hotlines or ethics lines, as they are known today, have become one of the main elements of the systems for managing, monitoring and ensuring compliance with the ethical dimensions underlying the different areas of human activity (García-Marzá, 2017; Calvo, 2020). This specific type of communication channel emerged in the professional world during the 1970s as a plausible tool for putting into practice the idea of whistleblowing.[2] This refers to the responsible actions of stakeholders inside the institution, organisation or company in monitoring[3] the behaviour of the institution or organisation and raising the alarm when there is evidence of destructive or immoral practices that could affect both the institution or organisation and society (Calvo, 2016).

---

[1] Data, from the Latin *datum*, means "what is given". Here, it is understood as the raw material from which information is formed. In other words, a piece of data is a symbolic (digital, spatial, alphabetical, etc.) representation of a qualitative or quantitative attribute or variable (about an organisation or event) which, when grouped and placed in order, generates a packet of information. A letter of the alphabet, for example, is data.

[2] The neologism "whistleblowing" emerged from the "call to responsibility" promulgated by the political activist Ralph Nader on 30 January 1971 at a congress on professional responsibility held in Washington DC (Boffey 1971: 549-551).

The term ethics hotline was used for the first time in the legal sphere, specifically by litigation specialists in the United States of America during the 1970s. The first record of its use and application dates from 1975, when the State Bar of California, in *State Bar of California Reports* no. 15 (1975, p. 65), announced that the Trial Lawyers' Department had set up an ethics hotline to improve compliance with its standards of conduct.

Although the concept was not yet widespread, the idea now formed part of the collective imagination of other professional spheres beyond the law. As mentioned in the study "Corporate Response to a New Environment" (van Dusen-Wishard, 1977), during the 1970s, a group of companies began designing and applying confidential channels for suggestions, queries, complaints and denunciations on ethical compliance for the first time: "Several leading companies, such as General Electric, Dow Chemical, IBM, and American Airlines, have established communication mechanisms which allow an employee to remain anonymous while challenging management decisions and practices," (Van Dusen-Wishard, 1977, p.230). Although the study does not specify the term used by these companies to refer to these compliance processes and/or channels, there is no doubt that the idea underlying them is the same one later linked to ethics hotlines.

Along the same lines, the study "Ethics. Knowing how to blow the whistle" (Perry, 1981, pp.56-61) published in the journal *IEEE spectrum* in September 1981, showed the emergence of whistle-blowers in the area of engineering during the 1970s in the United States of America, especially in sectors such as pharmaceuticals, nuclear energy and marine biology, as well as the value given to the whistle-blower and the action of whistleblowing by both the sector and society: "In fact, engineers who have taken critical stands publicly (…) sometimes find themselves honored as heroes of the profession. In cases where no one informs the public and flawed products lead to injuries, the public is quick to ask why someone didn't blow the whistle," (Perry, 1981, p.56). However, in these cases the alarm was raised or abuses reported mostly through the press or via notifications to the competent bodies among the public authorities. As the study stresses, during the 1970s few corporations tried to internally systemise the internal alarm and reporting process for possible unethical behaviour using some kind of confidential process or channel, although IBM once again stood out as a pioneer of this kind of action. Even in the study already mentioned (van Dusen-Wishard, 1977), the processes and/or channels were not linked to any specific term, despite their clear relationship with ethics hotlines.

---

[3] Here, "monitoring" is understood to be the action or effect of monitoring using different communication mechanisms so that the alarm can be raised if there are anomalies of an ethical nature in the institution, organisation or company and being involved in improving, correcting or eradicating such anomalies via suggestions or improvement proposals (Calvo, 2018).

> To enable engineers and others to give their best technical opinions even when the opinions are controversial, a small number of corporations have an internal mechanism that gives employees a fair hearing. Though the majority of cases handled under these systems are employee complaints and grievances, they can be and are used for ethical questions. The IBM Corp. is considered a pioneer in institutionalizing such procedures; its programs have been used as a model by other corporations. According to an IBM spokesman, under the company's Open Door policy, an employee with a complaint is urged to discuss it with the immediate manager, the manager's manager, or the site manager. Then, if the employee is not satisfied or chooses to skip the first step, he or she can get in touch with a general or regional manager and can even go further by writing to the chief executive officer and describing the situation or requesting a meeting. IBM also has what it calls a Speak Up program to deal confidentially with written complaints. It has handled 200,000 letters since its inception, and the company says it receives approximately 18,000 such contacts annually from IBM's 341,000 employees worldwide (Perry, 1981, pp.60-61).

Everything seems to indicate that general use of the term ethics lines in the professional sphere goes back to 1985, as shown in the studies "The Institutionalization of Organizational Ethics" (Sims, 1991) and "An Evaluation of the Ethics Program at General Dynamics" (Barker, 1994). The first of them, which emphasises how during the 1980s many companies had begun using mechanisms to ethically guide decision-making processes, such as codes of ethics and ethical training programmes, stresses that "General Dynamics has even gone to the point of implementing an 'ethics hot line' to aid employees when confronted with an ethical decision" (Sims, 1991, p.494). The second study confirms this statement and goes on to say that the idea of General Dynamics establishing an ethics hotline began in mid-1985 as a response to the serious accusations of fraud, deceit and extortion made against the company by the government of the United States of America (Baker, 1994).

Another example of how, during the 1980s, the design, application and implementation of whistleblowing processes and channels for ethical compliance, such as the use of the term ethics hotlines to define them, became widespread is its use in research. The first piece of information in this respect appeared in the *ABA Journal. The Lawyer's Magazine*, when in its July 1987 issue it informed its readers of the implementation of an ethics hotline by the Ethics Department at the ABA Center for Professional Responsibility. It explained that this was an ethics research assistance service whose main aim was to offer a tool for resolving queries about rules and standards, ethical opinions and other relevant issues for the ABA. The document also specified the ethics line communication channels: a telephone number, a postal address and a NET point (ABA, 1987, p.84).

As can be appreciated in these and other examples, since ethics lines emerged, their use and functions have expanded. They began as mere tools for raising the alarm and reporting irregularities, but other functions were quickly introduced, such as resolving queries about how to act ethically in specific cases. To some extent they became complementary to other tools – for example, their links with different types of ethics committee and spaces for face-to-face dialogue with experts on the subject. This made them channels for communication, innovation and participation suitable for involvement in recreating ethical ecosystems in the practical sphere.

Ethics lines have developed and considerably improved as communication channels to become mechanisms for innovation and participation through the incorporation of a) moral deliberation, by being complementary to ethics committees (Doudera 1983; White 2011); b) the committed participation of civil society in the processes of public scrutiny through non-financial reports and other communication mechanisms; c) opinions, suggestions and proposals to improve ethical systems, channels, codes and behaviour; and d) the compilation, processing and recording of best practices in terms of ethical behaviour and compliance (García-Marzá, 2017, 2019).

This development has made it possible to go beyond merely showing up dangerous or undesirable practices as an internal means of controlling irregularities and managing complaints. Much more ambitious goals can be set, such as a mechanism overseeing the good health – applicability and validity – of the code of ethics and ensuring improvements and compliance with the values, behaviour and commitments included in it. As a result, the ethics lines developed within ethics ecosystems offer the possibility of generating results linked to an increase in relevance, affinity, innovation, knowledge generation, conflict resolution and so on.

## 2.  TOWARDS THE DEVELOPMENT OF ETHICS LINES BASED ON DATA MINING

Nowadays there are designs for ethics lines that include all the functions, goals and expectations we have mentioned. A good example of this is the ethics line designed, developed and applied by García-Marzá together with his research team (García-Marzá, 2017, 2019; Calvo, 2018, 2020). This line, which has already been implemented and tested in institutions, organisations and businesses in sectors such as health, chemicals, education, tourism and ports,[4] is introduced directly in the code of ethics and conduct as a method of communication, participation and innovation to prevent incidents and improve compliance with the commitments made and the expected conduct. It is developed and implemented as a synergistic complement to the code of ethics and conduct, which offers the framework of moral values and standards that have to be followed; the ethics committee, which recreates the space for deliberation by internal and external stakeholders over the news received; the social responsibility or non-financial report, making it possible to publish the answers given to the news received to check whether or not there is an agreement with those affected; and the explainability report offering information on the design, application and use of digital techniques and technologies. In this way, the design of the ethics line consists of three parts:

1. Framework: understanding the codes, guidelines and standards making up the frame of reference for the ethics line. This mainly involves establishing a code of ethics and conduct publicly showing the values, commitment and conduct of the institution, organisation or business, as well as the national, international or sector guidelines that considered important to implement, such as the Universal Declaration of Human Rights; the relevant charters of rights and duties, and so on. This limits the type of notification that can be sent via the ethics line.

2. Structure: understanding the parts making up the line and the levels of application of the process. The most important parts are the code of ethics and conduct, the communication mechanisms, the ethics committee, the person responsible for the ethics line, the ethics line technician, the protocol for action, the record of notifications, the best practice archive, the results report, and so on. The levels of application of the process focus on how a person uses the ethics line to give notice of an issue they want dealt with [Level 1]; the receipt and technical analysis of notifications received for screening and allocation [Level 2]; dialogue on improvement proposals and best practice or, if concerning whistleblowing, consideration of the values at stake and possible courses of action to resolve or prevent conflicts [Level 3]; and the exploitation of information in order to prevent, eradicate or promote practices, processes and behaviour by the institution, organisation, company or associated professional [Level 4].

3. Protocol: includes the operation and procedures of the ethics line and its parts. The following should be particularly highlighted: the operation of the different communication channels making up the line, such as e-mail, telephone, websites, offices, QR codes, etc.; the competences and responsibilities of the person in charge the line, who above all should oversee the proper operation of the whole line, the confidentiality of the information and the protection of users; the tasks of the line technician who, among other things, must check the entry of notifications, determine their suitability and send them on wherever is appropriate; the application of the legal regulations such as the Data Protection Act or the Whistle-blower Protection Act; the composition, competences and operation of the ethics committee responsible for dialogue, deliberation and drawing up reports to provide responses or solutions for the notifications received; the process manual specifying the parties involved and the time for acknowledging receipt, investigating, and considering cases and the drawing up of response or solution reports; the penalty manual specifying types and degrees of breach and their consequences: warnings, beginning proceedings, attendance on training courses, the beginning of criminal proceedings, etc.; the function, goals and ethical audit procedure responsible for confirming the proper operation of the whole system and making a report on the results; the operation of the archive where the notifications, established procedures, responses given and solutions adopted are all recorded, etc.

---

[4] These ethics lines have been implemented by García-Marzá and his research team at the Universitat Jaume I (2018) in a health organisation (Unión de Mutuas, 2015); the chemical multinational UCE (UBE Corporation Europe, 2014); in the tourism sector in the Valencian region via Turisme Communitat Valenciana (2016-2019) and at the Port of Castelló (2019 and 2021). For further information in this respect, see the website of the *Ética práctica y Democracia* research group (2018).

An ethics line designed like the one proposed and developed by García-Marzá and his research group García-Marzá, 2017, 2019; Calvo, 2018, 2020) can, therefore, be a key element in managing, monitoring and compliance with ethics in different contexts of activity. Among other reasons, this is because the tool can help establish a diagnosis of the general rate of compliance with ethical standards in a sector, institution or organisation; detect or prevent bad practice or behaviour; receive suggestions and proposals for reviewing or improving the content and application of the code of ethics, standards of conduct or proceedings involved; detect conflicts, challenges and opportunities; monitor the gap between what is said and what is actually done, etc. Today, however, the digital transformation process and diverse and versatile techniques such as data mining make it possible to glimpse new paths and possibilities for development of ethics lines.

Data mining – a set of techniques for characterising, distinguishing, associating, classifying, predicting, grouping, analysing outliers (atypical values) and analysing

development – is a new level of the process of Knowledge Discovery in Databases (KDD).[5] However, its capacity to produce relevant information via analysis of large databases has made it now a core element of the process, because of its capacity to reveal patterns in a set of data.

In the development of ethics lines, association rules have become one of the data mining techniques arousing greatest interest. These are a subset of data mining which use automatic learning mechanisms to identify solid rules in large databases based on some interesting measures (Piatetsky-Shapiro, 1990). In other words, "(…) relationships within a set of particular transactions, items or attributes tending to occur together" (Amat-Rodrigo, 2018). In this respect, a critical and prudent application of this type of data mining technique offers notable possibilities for the development of ethics lines:

- Firstly, it makes it possible to scrutinise the information and carry out a situation diagnosis of the moral behaviour of the institution or organisation and its professionals, detecting anomalies, potential and/or weaknesses from an ethical point of view.

- Secondly, it makes it possible to extract relevant information to improve the understanding of the correct or incorrect behaviour of the institution or organisation and its professionals. For example, customising training programmes for a particular department, area or working group; the need to affect a particular aspect to promote or eradicate it; detecting failings in the system itself; failure to be specific enough about a particular value in the code of ethics, and so on.

- Thirdly, it makes it possible to identify indications of possible critical and/or post-conventional behaviour by analysing outliers that could affect the practical development of the institution or organisation through specific, customised actions.[6]

---

[5] Knowledge Discovery in Databases (KDD) is defined as: "(…) the nontrivial process of identifying valid, novel, potentially useful, and ultimately understandable patterns in data" (Fayyad et al., 1996, p.83).
[6] The outlier analysis requires non-digital variable analysis techniques which are therefore different to the ones used in this study. However, this study has considered incidents occurring least frequently as outliers.

Based on these ideas, an example of a model algorithm for extracting rules of association from the information collected and stored via an ethics line is proposed below. It must be said that the aim is not to offer answers or solutions to observable problems with this type of mathematical model. Its task in ethics lines is to make use of the data being generated and compiled in order to establish sets of items or attributes.[7] Based on patterns, these help those involved and/or affected to organise data; generate relevant, applicable information; and identify possible anomalies in the system and user behaviour to be reviewed, corrected, encouraged or expanded, as appropriate.[8]

### 3. A DESING AND SIMULATION OF THE ASSOCIATION RULES ALGORITHM FOR EXPLOITING DATA FROM THE ETHICS LINE

Exploiting the data compiled and generated by the ethics line allows, among other things, the detection of anomalies in the system and the extraction of behaviour patterns for action, encouragement or expansion, as appropriate. This requires at least five stages: *production* (generating data through the reports), data *preparation* (pre-processing the data for proper inclusion in the algorithm), *analysis* (study of the data collected using a proposed *Apriori algorithm*), *visualisation* (identifying the behaviour patterns underlying the data generated), and *action* (possible courses of action for solving the problems detected or correcting, encouraging or expanding the conduct observed) [Fig. 1].

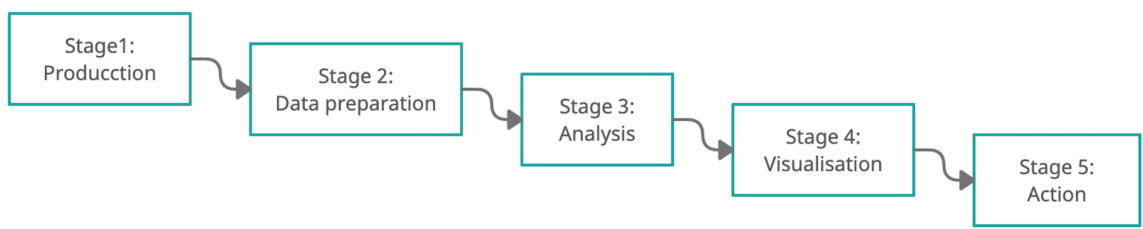

**Figure 1.** *Stages of the process of exploiting data from the ethics line*

---

[7] Here, we understand an attribute to be the properties or qualities of events represented symbolically.
[8] However, so as not to fall into the trap of positivism and scientificism, possible anomalies must be scrutinised and ultimately interpreted by those responsible for the line and/or those affected.

We will now describe a simulation of the application of this process carried out with synthetic data generated ad hoc for the purpose. The data is public and accessible via the *gitlab* web repository.[9]

a) *Stage 1. Production*

To obtain behavioural data via the ethics line, a Notification Form (proposal for improvement or best practice and alert or complaint about bad practice) has been created.[10] At Level 1 of the process of applying and implementing the ethics line, we find notification of good or bad ethical practices – particularly linked to compliance or non-compliance with the code of ethics and conduct. The screening and assignment of notifications at Level 2. In this respect, firstly, a preliminary questionnaire has been drawn up to extract data. It sets out some of the information that can be processed by the algorithm which can provide a diagnosis of the situation of the application of ethical values and other relevant issues: *tag, age, department, gender, specialisation, etc.*

In addition, to be able to analyse the data generated and/or collected on the notification forms, the dataset has been divided into bad and good practices. The process is therefore divided into two different runs of the algorithm:

- o Bad Practices: where a user makes a complaint or accuses another user, so two subjects are involved in the data processing;
- o Good Practices: where users highlight a good action or idea in general terms, so no second user is involved.

---

[9] All the data in the simulation is open access. It can be consulted at the website: https://gitlab.com/algorithms.ethicshotline/arulesethics/

[10] A second document has been developed for the extraction and collection of data for real cases with large databases. Level 3 of the process of applying and implementing the ethics line includes dialogue on proposals for improvement and best practices or, for complaints and alerts, dialogue and/or deliberation on the values at stake and possible courses of action to prevent or resolve conflict. To this end, *resolution report* templates have been developed to extract data that can be processed by the algorithm: the values at stake, courses of action, materiality, etc. However, this report has been excluded from the simulation because it requires techniques and algorithms different to the ones discussed in this study.

In addition, the data generated and collected through the notification reports has been structured based on two tables to make it more understandable [Fig.2]

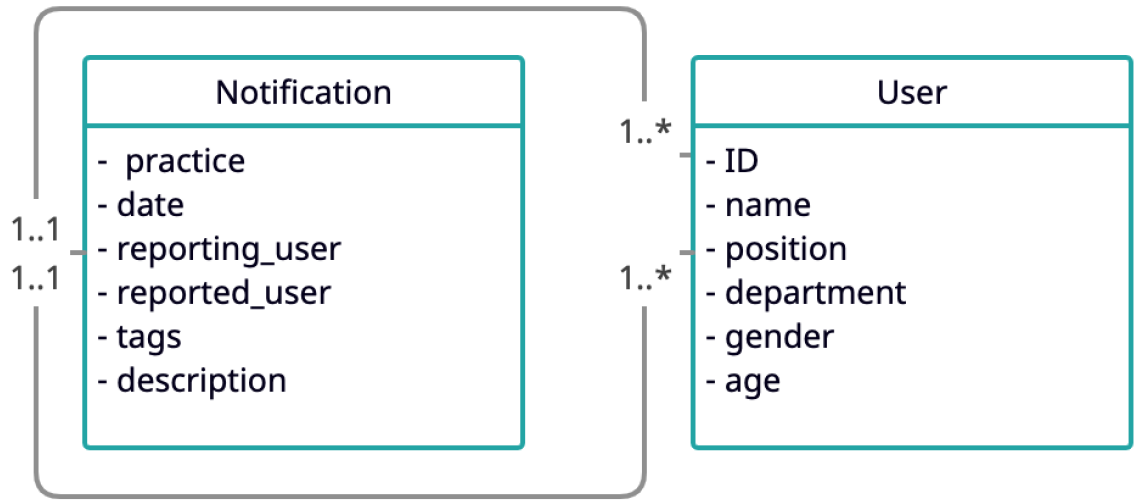

*Figure 2: Tables of data from the notification report*
***Source:*** *Own creation*

Part of the work of this study is to show an example of the application of an association rules algorithm with data from the following tables. A set of synthetic data has been used to create examples of this.[11] Here is one of them:

| ID | name | position | department | gender | age |
|---|---|---|---|---|---|
| u9 | Ramon Bilbao | employee | sales | m | 48 |
| u7 | Luisa Traver | manager | purchasing | f | 35 |
| u6 | Adriana Beltran | employee | production | f | 34 |
| u4 | Victor Regalado | employee | production | m | 28 |
| u26 | Miguel Suñer | employee | purchasing | m | 39 |
| u25 | Javier Caballero | employee | purchasing | m | 48 |
| u19 | Pedro Olet | employee | production | m | 21 |
| ... | ... | ... | ... | ... | ... |

---

[11] The data presented does not in any way correspond to reality. It has been generated *ad hoc* for the simulation described here, so any possible resemblance to reality should be treated strictly as coincidental.

| Practice | Date | Reporting user | Reported user | Tags | Description |
|---|---|---|---|---|---|
| b_p | 10/12/2012 | u1 | u0 | explainability report, sustainability | The HR department has drawn up an explainability report on the use and impact of digital devices |
| m_p | 1/2/2012 | u2 | u3 | conflict of interest, integrity | An administrative employee warns about a possible conflict of interest in the purchasing of office supplies. It has come to her attention that the company awarded the contract belongs to a close relative of the manager of the Purchasing Department |
| ... | ... | ... | ... | ... | ... |

***Table 2:*** *Notification table*

***Source:*** *Own creation*

b) *Stage 2. Data preparation*

Before moving on to the stage of applying the algorithm that finds the association rules, the data has to be prepared, eliminating duplication, data entry errors, missing values in some table fields, etc. This stage is essential in a processing environment with large data sets. No action should be taken on the data without good reason, as interference at this point may alter the results.

In this regard, the actions carried out at this stage are as follows:

- Removing the tag and description columns for processing in the association rules algorithm;

- Modifying the age column into five-year ranges for better processing in the association rules algorithm;

- Condensing the two tables into one data set to relate user and notification information;
- Dividing the data into two different sets, one for bad practices and one for good practices.

Bad practices: this type of practice involves both the reporting and reported parties (we will call this the *bad practices dataset*). The data will therefore take the following form [Table 3].

| practice | reporting user | reported user | reporting user's position | reporting user's department | reporting user's gender | reporting user's age | reported user's position | reported user's department | reported user's gender | reported user's age |
|---|---|---|---|---|---|---|---|---|---|---|
| bad_practice | u2 | u3 | employee | production | f | 30-34 | manager | purchasing | m | 55-59 |
| bad_practice | u4 | u5 | employee | production | m | 25-29 | business person | not_applicable | f | 55-59 |
| bad_practice | u6 | u0 | employee | production | f | 30-34 | not_applicable | not_applicable | not_applicable | 35-39 |
| ... | ... | ... | ... | ... | ... | ... | ... | ... | ... | ... |

**Table 3:** *Good practices table*
**Source:** *Own creation*

Good practices: this type of practice involves only the reporting user, so there are no columns for the reported user (we will call this set the *good practices dataset*). The data will take the following form [Table 4].

| practice | reporting user | reporting user's position | reporting user's department | reporting user's gender | reporting user's age |
|---|---|---|---|---|---|
| good_practice | u1 | employee | HR | m | 35-40 |

| good_practice | u4 | employee | production | m | 25-29 |
| good_practice | u13 | employee | sales | m | 35-40 |
| ... | ... | ... | ... | ... | ... |

c) *Stage 3. Analysis*

Level 4 of the process of applying and implementing the ethics hotline involves analysing the data collected and generated by the ethics hotline through the *notification forms*. For this purpose, an algorithm model like the one designed by Rakseh Agrawal and Ramakrisnnan Srikant in "Fast Algorithms for Mining Association Rules in Large Databases" (1994, pp.487-499) – known as *Apriori* – can be extremely useful. This is a mathematical model traditionally used in data mining which crawls large databases to efficiently locate sets of recurring items or attributes in order to obtain behavioural patterns from them [Fig. 3].

```
procedure LargeItemsets
begin
  let Large set L = ∅;
  let Frontier set F = {∅};

  while F ≠ ∅ do begin

    −− make a pass over the database
    let Candidate set C = ∅;
    forall database tuples t do
      forall itemsets f in F do
        if t contains f then begin
          let C_f = candidate itemsets that are extensions
                of f and contained in t; −− see Section 3.2
          forall itemsets c_f in C_f do
            if c_f ∈ C then
              c_f.count = c_f.count + 1;
            else begin
              c_f.count = 0;
              C = C + c_f;
            end
        end

    −− consolidate
    let F = ∅;
    forall itemsets c in C do begin
      if count(c)/dbsize > minsupport then
        L = L + c;
      if c should be used as a frontier −− see Section 3.3
        in the next pass then
          F = F + c;
    end
  end
end
```

*Figure 3:* *Pseudocode of the* Apriori *algorithm*
*Source:* *Agrawal and Srikant (1994: 487-499).*

An analysis model can be developed based on the *Apriori algorithm* designed by Agrawal and Srikant (1994, pp. 487-499), to extract behavioural patterns from the data collected and/or generated by the ethics line. This will significantly increase self-knowledge about the operation of the system, recurrent patterns and behaviours and the level to which the values from the code of ethics are being assumed by internal stakeholders.

a. Problem definition

Following the original definition provided by Agrawal, Imielinski & Swami in "Mining Association Rules Between Sets of Items in Large Databases" (1993, pp.207-216), the association rule mining problem is defined as follows: $I = I_1, I_2, \ldots, I_f$ is a set of binary attributes, called items. $T$ is a transaction database, where each transaction *(T)* has its own identifier (ID) and contains a subset of item (II) (Agrawal, Imielinski & Swami, 1993, p.208).

Thus, an association rule is defined as an implication of the form $X \Rightarrow Y$, where $X, Y \subseteq I \land X \cap Y = \varnothing$. The sets of items $X$ and $Y$ are called respectively "antecedent" (left) and "consequent" (right) (Agrawal, Imielinski & Swami, 1993, p.208).

b. Measures to be taken into account

In order to use association rules on the information from the ethics line, we must remember that constraints must be used on various important and interesting measures in order to select the relevant rules from the set of all possible rules. The best-known limitations are the support, confidence and lift thresholds.

- o Support: indicates the frequency with which the set of elements appears in the data set. As for the support of $X$ with respect to $T$, it is defined as the proportion of transitions $t$ in the data set containing the set of elements $X$.

$$supp(X) = \frac{|\{t \in T; X \subseteq t\}|}{|T|}$$

- o Confidence: indicates the frequency with which the rule has been found to be true. The confidence value of a rule, $X \Rightarrow Y$, with respect to a set of transactions $T$, is the proportion of the transitions containing $X$ that also contain $Y$. It is defined as:

$$conf(X \Rightarrow Y) = \frac{supp(X \cup Y)}{supp(Y)}$$

- o Correlation: this indicates the lift or correlation of a rule, which is defined based on the following formula:

$$lift(X \Rightarrow Y) = \frac{supp(X \cup Y)}{supp(X) \times supp(Y)}$$

If the relationship between the observed and expected support is independent, the following possibilities arise:

- If the rule had a correlation of 1, it would imply that the probability of the occurrence of the antecedent and that of the consequent are independent of one another. When two events are independent of one another no rule involving the two events can be established.

- If the correlation is >1, we know the degree to which those two occurrences are dependent on one another. That makes these rules potentially useful for predicting the consequent in future data sets.

- If the correlation is <1, we know that the elements replace one other. This means the presence of one element has a negative effect on the presence of the other element and vice vera.

    c.   Running the algorithm

To run the Apriori algorithm with synthetic data, we will use the [apriori](#)[12] function from the library *Arules* [version 1.6-4](#)[13]. This function is implemented in [r language](#)[14], one of the most common languages, together with Python, for running machine learning algorithms.

Running the apriori function requires a series of parameters requiring us to select some of the rules obtained in the function output, filtering out the rules we do not consider interesting. When running the algorithm within the project we consider the following:

- support = 0.4 -> minimum support of the rule to be considered (appearing in at least 40% of the total);

- confidence = 0.8 -> minimum confidence of the rule for it to be considered (that when the antecedent appears, the consequent appears in at least 80% of cases);
- minlen = 2-> minimum number of rules that must appear as the sum of the antecedent and the consequent (there must be at least one antecedent and one consequent).

The algorithm will normally run in the following form:

```
rules <- apriori(data,
               parameter = list(supp = 0.4, conf =0.8, minlen = 2))
```

The algorithm is applied to the two data sets, the *bad practice dataset* and the *good practice dataset*.

*d) Stage 4. Visualisation*

Using the *Apriori algorithm* model proposed in *Stage 3. Analysis,* for the processing of the data generated and collected by the ethics hotline using the different forms and reports, can have an impact on a new level of application and implementation of the ethics hotline. This Level 5 would consist of converting the data into relevant information by using, for example, association rules – behavioural patterns. These would make it possible to visualise possible opportunities, potentialities and weaknesses – in other words, anomalies in the system or recurring behaviours – which, depending on how they are assessed, can be promoted or eradicated.

The simulation results, which are divided between good and bad practices, have yielded the following graphs and results:

- Application of the Apriori function to the bad practice dataset [table 5].

| lhs | | rhs | support | confidence | lift | count |
|---|---|---|---|---|---|---|
| {age_reporting_user=30-34} | = | {department_reporting_user=produc | 0.4286 | 0.8824 | 1.5837 | 30 |
| | > | tion} | | | | |
| {age_reporting_user=30-34} | => | {position_reported_user=employee} | 0.4143 | 0.8529 | 1.4926 | 29 |
| {age_reporting_user=30-34} | => | {gender_reporting_user=f} | 0.4 | 0.8235 | 1.4412 | 28 |
| {age_reporting_user=30-34} | => | {position_reporting_user=employee} | 0.4857 | 1 | 1.129 | 34 |
| {department_reporting_user=production} | => | {position_reporting_user=employee} | 0.5571 | 1 | 1.129 | 39 |
| {position_reported_user=employee} | => | {position_reporting_user=employee} | 0.5571 | 0.975 | 1.1008 | 39 |
| {gender_reported_user=m} | => | {position_reporting_user=employee} | 0.4 | 0.9333 | 1.0538 | 28 |
| {gender_reporting_user=f} | => | {position_reporting_user=employee} | 0.5 | 0.875 | 0.9879 | 35 |

- 

- ***Table 5:*** *Bad practice table dataset*
- ***Source:*** *Own creation*

- 

The rules obtained have been ordered by their lift (correlation between them). As the function parameters have restricted its output to the rules that we find interesting, we will consider all of them valid. It can therefore be stated that the following rules apply for the simulated data set:

- When the reporting user is aged between 30 and 34, the reporting department is production;
- When the reporting user is aged between 30 and 34, the reported user is an employee;
- When the reporting user is aged between 30 and 34, their gender is female;
- When the reporting user is aged between 30 and 34 their position is employee;

- When the reporting user's department is production, their position is employee;

- When the reporting user is an employee, their position is employee;

- When the reported user's gender is male, the reporting user is an employee;

- When the reporting user's gender is female, their position is employee [Fig. 4]

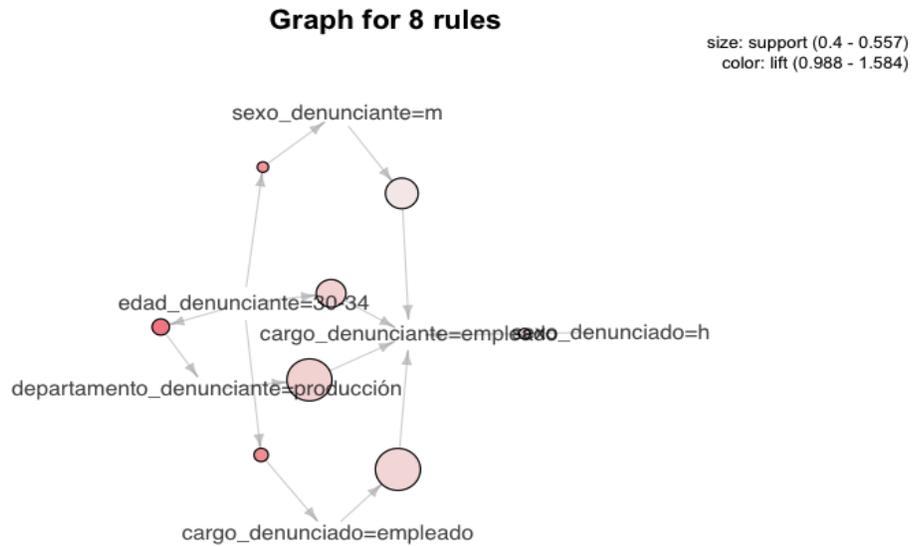

*Figure 4.* Ethics line association rule

**Source:** https://gitlab.com/algorithms.ethicshotline/arulesethics/

- Applying the Apriori function to the *good practices dataset* [Table 6]

-

| - lhs | | rhs | support | confidence | lift | count |
|---|---|---|---|---|---|---|
| {department_reporting_user=hr} | => | {position_reporting_user=employee} | 0.666666666666667 | 1 | 1 | 12 |
| {gender_reporting_user=f} | => | {position_reporting_user=employee} | 0.666666666666667 | 1 | 1 | 12 |

**Table 6:** *Good practices table dataset*

**Source:** *Own creation*

There are only two rules that apply to good practices. It should be borne in mind that this type of algorithm is created for large data sets, so in small sets not many rules will be extracted. This may be an indication that more data or greater effort is needed to collect the data required.

The rules that apply for this simulated data set are:

- When the announcer's department is HR, the announcer's position is employee;
- When the announcer's gender is female, their position is employee [Fig. 5].

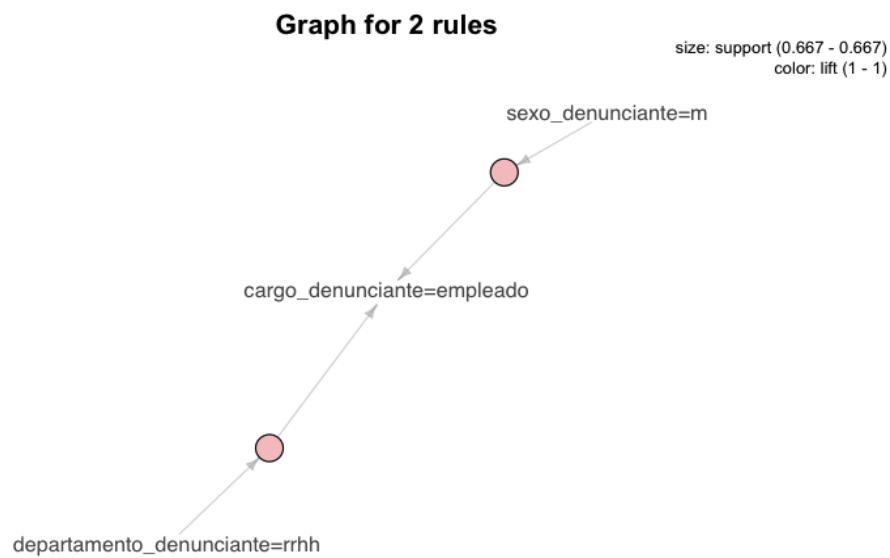

**Figure 5:** *Ethics line association rule*

**Source:** *https://gitlab.com/algorithms.ethicshotline/arulesethics/.*

The simulation of the application of the *mathematical* model proposed in *Stage 3* on synthetic data from the ethics line makes it possible to see that, through the correlations generated (information), it is possible to identify current or potential strengths, weaknesses, opportunities and threats, as well as possible courses of action for eradicating, developing or promoting them (knowledge). Here are some examples:

First, the attributes *tags* and *description* [the specific issues and reasons for notification] may reveal emotional, conflict-related, behavioural, performance-related, reactive, proactive, propositional, conventional, critical and/or punitive factors that can be adequately managed via dialogue or deliberation processes. For example, the simulation has shown that a good number of complaints concern abusive practices and warnings of corrupt practices. This may mean there is an anomaly in the organisational culture linked to two fundamental aspects of excellent character: equality and integrity.[15]

Secondly, the attribute *age* [the age of the informant and/or the subject(s) about whom they are reporting] can help identify *generational* patterns and behaviours for action, either to eradicate, promote or expand them, through specific training programs. For example, the simulation showed a possible relationship between bad practice and men aged between 30 and 34.

Thirdly, the attribute *department* [unit or division to which the people involved belong] and their position [post held by the people involved] can help show up corporate patterns and behaviours to be eradicated, promoted or expanded by redesigning them or creating and/or promoting the appropriate culture. For example, the simulation showed there is a certain tendency towards peers reporting one another. This could be due to a fear of dismissal when reporting people in senior positions. It also showed that notifications of good practices are mostly concentrated in the human resources department. This may be due to the specific training courses on particular aspects of the code of ethics, such as equality, integrity, etc., which this department, unlike the others, has been holding.

Fourthly, the attribute *gender* [sex of the people involved], may be showing a positive or negative tendency linked to the *gender role* assumed by the people concerned. For example, the simulation has shown greater involvement by women in pointing out good practices and a greater accumulation of complaints of bad practice about men behaving inappropriately. This can be interpreted as a greater commitment by women to the good health of the organisation, as well as a greater tendency by men to engage in inappropriate practices.

---

[15] The simulation has shown that the synthetic data set used is too poor to generate relevant information on these two attributes. For this reason, it was not considered appropriate to provide examples of the simulation results.

Fifthly, the attribute *practice* [whether good or bad practice is reported] can provide information about the degree of corporate (general or departmental) commitment and/or ownership of the values and behaviours in the code of ethics, as well as the degree of influence of other attributes – generational, industry-related, corporate, social, departmental and gender-related. For example, receiving notifications about the periodic production of a report on the collection, processing and use of bulk data by an organisation, department or working group can show information about the level of ownership and commitment associated with the value of transparency. However, if there are significant differences in the adoption of these practices between different areas or professional groups within the corporation, this could indicate the influence of attributes such as the department or position, for example.

However, the greatest potential for establishing association rules lies in revealing complex correlations (patterns) established on the basis of multiple attributes. For example, in a specific case where a pattern is revealed between *reason* (bad practice), age (30-35), gender (female), *department* (finance) and position (manager), it is possible to determine how and to what degree certain behaviours are affected by generational, corporate, industry-related, departmental, professional, behavioural, axiological and/or gender-related factors. This allows actions to be adjusted, guiding the culture, communication and training of the institution, organisation or company. However, as mentioned above, the proposed algorithm is applied to large quantities of data.

*e) Stage 5. Action*

The final stage of the data generation and exploitation process, which in this case corresponds to Level 6 in the application and implementation of the ethics line, concerns the possible courses of action and the different instruments for its practical application to correct anomalies in the system or to encourage, eradicate or expand the behaviour patterns detected.

The following table shows the different attributes used to simulate the application of the algorithm, the attribution of different associated factors and some of the possible actions that could be taken to encourage, eradicate or expand the behaviour patterns [Table 7].

| Attribute | Factor | Actions |
|---|---|---|
| Tags | Conflict-related Behavioural Performance-related Propositional Preventive Punitive Critical | Dialogue and deliberation processes, preparation of reports (on improvement, best practices, conflict resolution, etc.), public records, preparation of specific guidelines, etc. |
| Description | Conflict-related Behavioural Performance-related Propositional Preventive Punitive Critical | Dialogue and deliberation processes, preparation of reports (on improvement, best practices, conflict resolution, etc.), public records, preparation of specific guidelines, etc. |
| Age | Generational | Multi-topic or specific training |
| Department | Corporate | Redesign, culture, specific training |
| Position | Sector-specific | Culture, specific training, character |
| Gender | Gender | Personal, departmental or multi-topic training, code of conduct, equality plan |
| Good_practice | Axiological, behavioural, procedural | Records, visibility, promotion, encouragement |

| Bad_practice | Axiological, behavioural, procedural | Records, visibility, promotion, encouragement |
|---|---|---|

This *Stage 5. Action* draws significantly on all the knowledge gained from the four previous phases, but its implementation is entirely the responsibility of the ethics committee. The committee analyses the results and, after dialogue and deliberation, agrees the measures to be taken, such as revising the code of ethics, deliberating on a case of malpractice or exerting influence over some type of specific or more general training (García-Marzá, 2004, 2017; Calvo, 2021).

### 4. Conclusions and limitations of the study

New digital technologies linked to the general digital transformation open up new and complex challenges for ethics, but there are also opportunities to develop and apply them in the practical field. For example, data mining techniques such as *Arules* make it possible to exploit data linked to the actual behaviour of the institution, organisation or company. Specifically, in this study we have tried to show their potential application to ethics lines, as well as outlining possibilities and guidelines for their design and practical application, with a simulation of the application of the *Apriori algorithm* model.

However, there are still many rough edges to be polished and limitations to be overcome before the idea can be properly implemented. These include finding out how to improve the analysis and interpretation of anomalous behaviour (outliers) to generate relevant information on fraud, anomalies, trends and, above all, post-conventional criticisms to allow us to move towards a fair, congratulatory horizon for action.

The limitations of the simulation described must also be taken into account. Although this simulation provides an approach to the impacts that can be created using algorithms such as Apriori and data mining techniques like association rules, it would be necessary to experiment with the model on large data sets to test its reliability and improve the results. The next step would therefore consist of carrying out a simulation on an ethics line that would generate and collect large quantities of data. This would make it possible to adjust the proposed model to make better use of the data and results.